\documentclass{article}
\usepackage{spconf,amsmath,graphicx, xcolor, amssymb}
\usepackage{url}
\usepackage{subcaption}

\def\x{{\mathbf x}}

\title{Fast PSF Synthesis with Defocused and Spherical Aberration}
\name{Nicholas Ganino and Qi Guo}
\address{Elmore Family School of Electrical and Computer Engineering \\
Purdue University \\
\url{qiguo@purdue.edu}
}
\begin{document}
%
\maketitle
\begin{abstract}
Accurately estimating the point spread function (PSF) of an optical system requires solving free-space wave propagation, which entails evaluating a diffraction integral. This integral is traditionally computed numerically using Fast Fourier Transform (FFT) or Hankel Transform, as it lacks a closed-form solution. We show that, under defocus and spherical aberration, the diffraction integral admits an approximate closed-form solution by combining a piecewise Bessel approximation with Gaussian-type integrals. Based on this result, we develop a fast wave-based PSF simulator with linear complexity in the radial resolution. The proposed, un-optimized simulator achieves up to a 2× speedup over Hankel-based integration and a 4× speedup over FFT while closely matching wave-optical PSFs, enabling efficient large-scale depth-of-field synthesis.
\end{abstract}
\section{Introduction}
\label{sec:intro}

\begin{figure}[!t]
    \centering
    \includegraphics[width=0.9\linewidth]{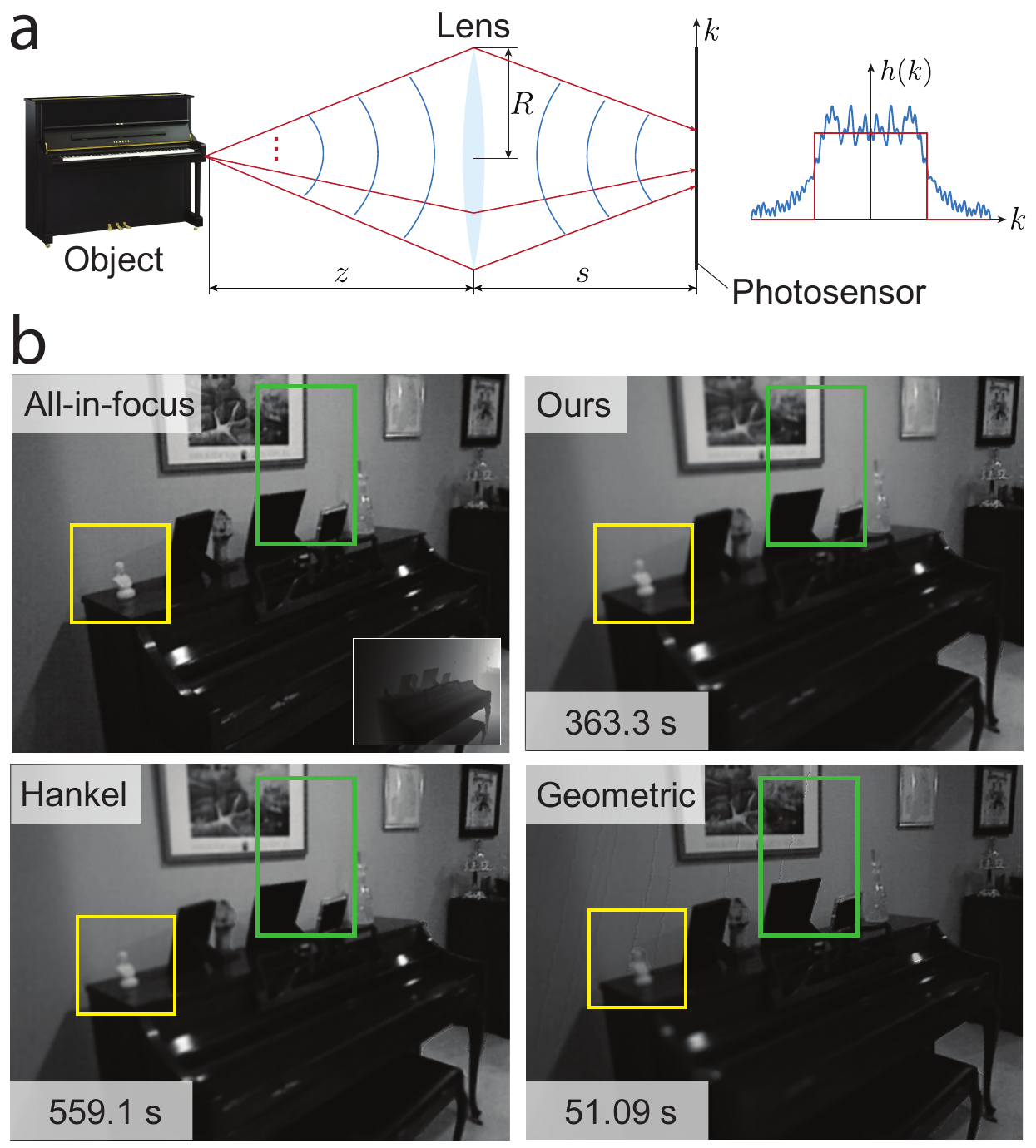}
    \caption{Overview. (a) Wave-based (blue) and geometric-based (red) PSF synthesis. Wave-based simulation accurately models diffraction at high computational cost, while geometric simulation is fast but inaccurate. (b) DoF rendering via per-pixel PSF synthesis according to the provided depth map (inset). Our simulator achieves substantially lower runtime than the Hankel-based wave simulator used in recent work~\cite{froch2025beating}, while keeping high fidelity. Boxes highlight discretization artifacts in geometric rendering.}
    \label{fig:teaser}
\end{figure}

Efficient and accurate synthesis of depth-of-field (DoF) images is critical for a wide range of computer vision and graphics applications, including depth from defocus~\cite{luo2025depth,xu2025blurry,wu2019phasecam3d}, image restoration~\cite{tsai2022stripformer,kong2023efficient,jiji2024extended}, and rendering~\cite{barsky2008algorithms, krivanek2003fast}. Despite its importance, DoF synthesis remains challenging because it requires evaluating point spread functions (PSFs) at a densely sampled set of image locations. Geometric PSF simulators~\cite{luo2025depth} are computationally efficient, but they fail to capture diffraction effects and often suffer from discretization artifacts, especially near focus (Fig.~\ref{fig:teaser}). In contrast, wave-based PSF simulators~\cite{sitzmann2018end} provide high physical accuracy but incur substantial computational cost due to the numerical evaluation of diffraction integrals. These limitations motivate the need for a PSF simulator that is both fast and accurate, enabling practical DoF image synthesis at scale.

This paper proposes a fast simulator that approximates the diffraction integral under defocus and spherical aberration in closed form. By avoiding expensive numerical integration, the proposed approach substantially reduces runtime while retaining diffraction effects, compared to wave-based simulators commonly used in recent work~\cite{froch2025beating, tseng2021neural, chan2023computational}. As shown in Fig.~\ref{fig:teaser}, our method produces visually accurate DoF images with a clear computational advantage over existing wave-based approaches, while also achieving higher accuracy than geometric simulation.

Existing wave-based methods follow the classic approach of Goodman~\cite{goodman1996fourier}, which typically evaluate the diffraction integral numerically using the Fast Fourier Transform (FFT). When the wavefront is radially symmetric, this evaluation can be reduced from two dimensions to one dimension and solved via the Hankel transform. Further simplification, however, is challenging due to the lack of closed-form solutions for diffraction integrals with general wavefront phase profiles. There has been work that combines geometric optics and wave optics to efficiently model aberrations in optical assemblies~\cite{ho2025differentiable, ren2025successive}. However, these models still require solving diffraction integrals within frameworks. Researchers have also trained neural networks to implicitly represent PSFs as a function of depth, incident angle, etc., but training these models still requires data generated from solving diffraction integrals~\cite{tseng2021differentiable, lin2025learning, mao2021accelerating}.  In this work, we directly tackle the simplification of the diffraction integral, which we have not seen in prior works. We show that when the wavefront is parameterized by Seidel coefficients consisting only of defocus and spherical aberration, the diffraction integral admits an accurate approximate closed-form solution. 

We derive this solution in detail, implement a fast PSF simulator based on it, and analyze its computational efficiency and accuracy relative to widely used FFT- and Hankel-based simulators. Experimental results demonstrate a $2\times$ speedup over Hankel-based methods and a $4\times$ speedup over FFT-based methods, while maintaining high fidelity in the synthesized PSF shapes. The proposed simulator is complementary to existing FFT- and Hankel-based approaches, trading generality for a significant gain in computational efficiency.

The main contributions of this paper are:
\begin{itemize}
\item An approximate closed-form solution to the diffraction integral under defocus and spherical aberration;
\item A fast PSF simulator based on this solution;
\item A comprehensive comparison with existing PSF simulation methods in terms of accuracy and runtime.
\end{itemize}
The PSF simulator based on this paper can be downloaded at: \url{https://hankel.qiguo.org}.

\section{Principle}

Consider an on-axis point source located at distance $z$ from a single-lens camera, emitting a spherical wavefront with wavelength $\lambda$, as illustrated in Fig.~\ref{fig:teaser}a. The lens is assumed to transmit all incident light within a circular aperture of radius $R$. Under these assumptions, the resulting point spread function (PSF) $h$ on the photosensor is radially symmetric and can be expressed via the well-known Hankel transform
\begin{equation}
    h(k) = \left|2\pi \int_{0}^{R} 
    P(r)
    J_0(2\pi k r)\, r\, dr\right|^2.
    \label{eq:hankel}
\end{equation}
The variables $r$ and $k$ denote the radial coordinate at the aperture and sensor plane, respectively, $J_0(\cdot)$ is the zeroth-order Bessel function of the first kind, and the pupil function $P(r)$ is given by: 
\begin{equation}
    P(r) = \exp\!\left(j 2 C_d\, r^2/R^2\right),
\end{equation}
where $C_d$ is the defocus coefficient that depends on how far an object is from the focusing distance $z_f$  \cite{wyant1992basic}:
$$
C_d = \frac{\pi (z_f - z) R^2}{2\lambda \cdot z \cdot z_f}
$$

Evaluating~\eqref{eq:hankel} typically requires either direct numerical integration, which can be formulated as a matrix multiplication with $O(N^2)$ complexity, or a Fast Hankel Transform (FHT) with $O(N\log N)$ complexity~\cite{hamilton2000uncorrelated}, where $N$ denotes the number of radial samples of the PSF.

\begin{figure}[htbp]
    \centering
    \includegraphics[width=\linewidth]{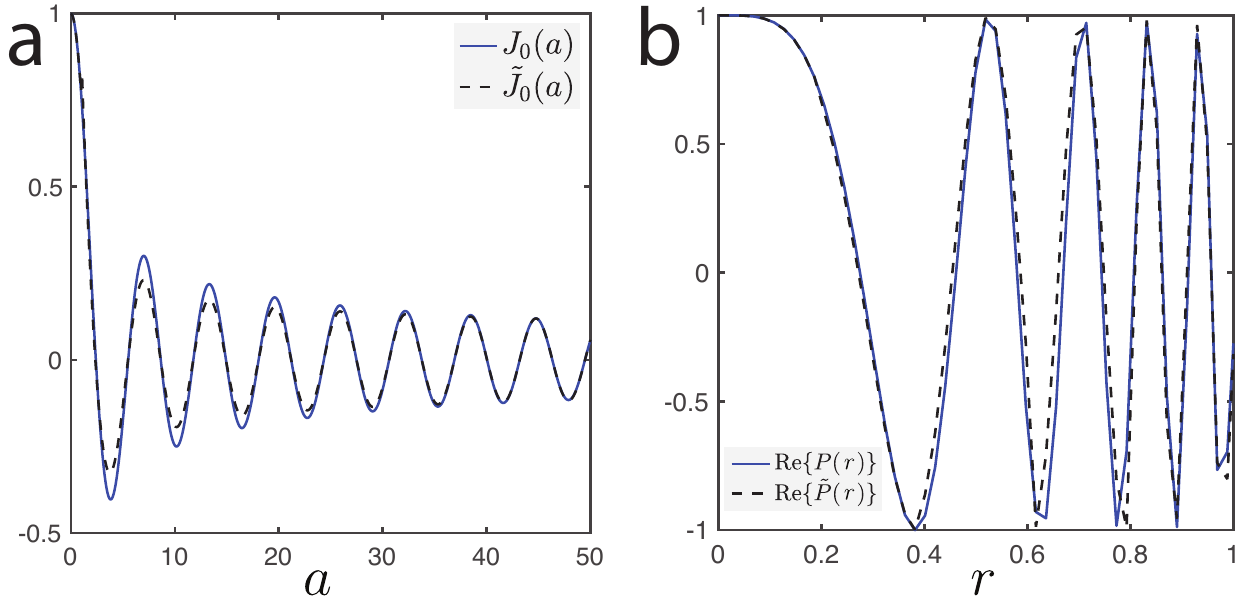}
    \caption{(a) Zeroth-order Bessel function of the first kind $J_0(a)$ and its approximation $\tilde{J}_0(a)$, with $\alpha =100$.
(b) Pupil function $P(r)$ and its approximation $\tilde{P}(r)$ with $C_d = 10$, $C_s = 10$, and $M=5$ partitions. Both approximations closely match their corresponding ground-truth functions.} 
    \label{fig:J0_Pupil}
\end{figure}

The key contribution of this work is an analytic approximation of the Bessel function $J_0(a)$ that enables a closed-form evaluation of the Hankel integral. Specifically, we adopt the following piecewise approximation:
\begin{equation}
\label{eq:J0_piecewise}
\tilde{J}_0(a) =
\begin{cases}
f(a), & a \leq 1, \\[4pt]
g(a), & \text{otherwise},
\end{cases}
\end{equation}
where
\begin{equation*}
    \begin{aligned}
    f(a) &= 1 - \frac{a^2}{4} + \frac{a^4}{64}, \\
    g(a) &= \sqrt{\frac{2}{\pi}}
    \left(
        \frac{3}{2}\alpha^{-\tfrac{1}{2}}
        - \frac{a}{2} \alpha^{-\tfrac{3}{2}}
        + \frac{1}{a}
    \right)
    \cos\!\left(a - \tfrac{\pi}{4}\right).
\end{aligned}
\end{equation*}
Here, $\alpha$ is a tunable hyperparameter controlling the operating point of the approximation. As shown in Fig.~\ref{fig:J0_Pupil}a, the proposed approximation $\tilde{J}_0(a)$ closely matches the true Bessel function $J_0(a)$ over the entire domain for representative values of the variable $a$.

Substituting $\tilde{J}_0$ into~\eqref{eq:hankel} yields a decomposition of the PSF into two integrals,
\begin{equation}
\label{eq:full_integrand}
\begin{aligned}
h(k) \approx 4\pi^2 \Bigg|
&\int_{0}^{\tfrac{1}{2\pi k}}
f(2\pi kr)\,
\exp\!\left(j 2 C_d\, r^2/R^2\right) r\, dr \\
+&\int_{\tfrac{1}{2\pi k}}^{R}
g(2\pi kr)\,
\exp\!\left(j 2 C_d\, r^2/R^2\right) r\, dr
\Bigg|^2.
\end{aligned}
\end{equation}
Both integrals admit closed-form solutions by reducing to Gaussian-type integrals of the form~\cite{abramowitz1964handbook}:
\begin{multline}
\label{AS_7.4.32}
\int \exp\!\left[-(\nu x^2 + 2 \gamma x + \epsilon)\right] dx = \\
\frac{1}{2}\sqrt{\frac{\pi}{\nu}}
\exp\!\left(\frac{\gamma^2 - \nu \epsilon}{\nu}\right)
\operatorname{erf}\!\left(\sqrt{\nu}\,x + \frac{\gamma}{\sqrt{\nu}}\right)
+ \text{constant}.
\end{multline}

As a result, the PSF can be evaluated analytically with only $O(N)$ computational complexity, representing a substantial reduction compared to existing numerical and FHT-based approaches. For brevity, we omit the explicit closed-form expression in the main text and refer the reader to the Appendix for the full derivation.

\paragraph*{Extension to Spherical Aberration.} When both defocus and spherical aberrations are present, the pupil function can be written as
\begin{equation}
    P(r)=\exp\!\left[j\left(\frac{2 C_d r^2}{R^2} + \frac{C_s r^4}{R^4}\right)\right],
\end{equation}
where $C_s$ denotes the spherical aberration coefficient. The presence of the quartic term $r^4$ in the phase prevents the reduction of the Hankel integral to closed form using the previous approach.

To address this difficulty, we approximate the pupil function using a piecewise quadratic phase model. Specifically, we partition the pupil into $M$ radial intervals $[r_i, r_{i+1})$, within each of which the phase is locally approximated by a quadratic function:
\begin{equation}
\label{eq:pupil_piecewise}
    \tilde{P}(r) = 
    \begin{cases}
        \exp\!\left[j\alpha_1 r^2\right], & 0 \leq r < r_1, \\[2pt]
        \exp\!\left[j\alpha_2 r^2\right], & r_1 \leq r < r_2, \\[2pt]
        \vdots & \\[2pt]
        \exp\!\left[j\alpha_M r^2\right], & r_{M-1} \leq r < r_M,
    \end{cases}
\end{equation}
where the coefficients $\{\alpha_i\}$ are chosen to best approximate the original quartic phase within each interval. As illustrated in Fig.~\ref{fig:J0_Pupil}b, the resulting piecewise approximation closely resembles the true pupil function.

With this approximation, each segment reduces to the same form as the purely defocus case in~Eq.~\ref{eq:full_integrand}, allowing the PSF contribution of each interval to be evaluated analytically using Gaussian integrals. The overall PSF is then obtained by summing the contributions across all segments.

\section{Results}
\label{sec:results}

This section analyzes the speed and accuracy trade-off of the proposed closed-form solution of defocused and spherically-aberrated PSFs. We chose the following methods as baselines, which have been frequently used in recent computational imaging works~\cite{hazineh2022d, froch2025beating, luo2025depth}. First, we evaluate Hankel transform (Eq.~\ref{eq:hankel}) via matrix multiplication and then interpolate the radial PSF $h(k)$ to the two-dimensional PSF $h(\x)$:
\begin{align}
    h(\x) \equiv  h(k=\lVert\x\rVert).
\end{align}
We did not utilize FHT, as the disk pupil function violates the smoothness constraint that FHT requires and the calculated PSF significantly deviates from the plain Hankel transform. We implemented Hankel transform in Python scripts. Second, we directly calculate the 2D PSFs according to:
\begin{align}
    h(\x) \propto \left|\mathcal{F} \left[ (\lVert\x\rVert < R) P(\lVert\x\rVert)\right]\right|^2.
    \label{eq:2d_psf}
\end{align}
via Fast Fourier Transform (FFT) by calling Scipy's fft function. 
Third, we use the simple geometric optics implemented in Python, where the PSF is a scaling of the pupil under the thin-lens model~\cite{luo2025depth}:
\begin{align}
    h(\x) = \frac{1}{\sigma^2} \left(\frac{\lVert \x \rVert}{\sigma} < 1\right), \text{where }\sigma = Rs\left(\frac{1}{z} - \frac{1}{z_f}\right),
\end{align}
where $s$ and $z_f$ are the sensor and focal distances, and $z$ is the target distance as in Fig.~\ref{fig:teaser}a. The defocus parameter $C_d$ can be similarly calculated as~\cite{wyant1992basic}:
\begin{align}
C_d = \frac{\pi (z_f - z) R^2}{2\lambda \cdot z \cdot z_f}
\end{align}
For simplicity, we refer to these baseline PSF evaluation methods as \textit{Hankel}, \textit{FFT}, and \textit{Geometric}. Hankel and FFT are wave-based, thus their outputs are consistent with each other, and can accurately render the diffraction effects in the PSFs, while Geometric cannot. We consider the output from Hankel and FFT to be the true PSFs in our evaluation. For our method, we compute the radial PSF $h(k)$ using the closed-form solution implemented in Python scripts and then interpolate it to 2D, similar to Hankel. 



\paragraph*{Range of realistic $C_d$ and $C_s$.} For the defocus parameter $C_d$, we set it to be smaller than $10$ as larger PSFs will likely blur out textures in images. We surveyed stock lenses sold by major optical elements suppliers, and found that the spherical parameter $C_s$ of typical lenses ranges from $0$ to $10$. 

\begin{figure}[!t]
    \centering
    \includegraphics[width=\linewidth]{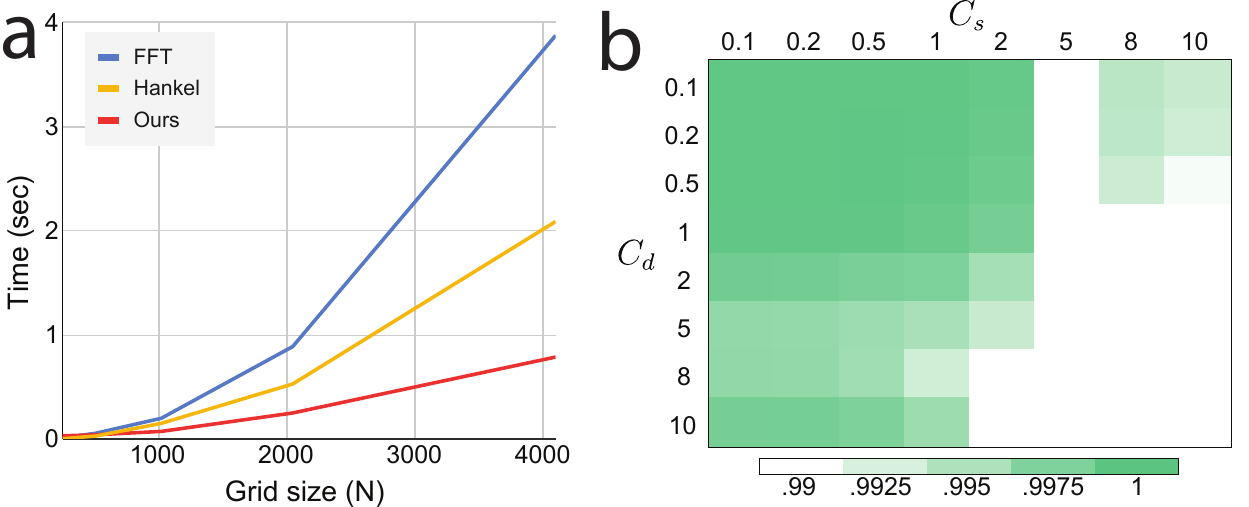}
    \caption{Quantitative comparison of PSF synthesis runtime and accuracy. (a) Runtime for generating a 2D PSF across varying spatial resolutions $N$. Our method is consistently 2× faster than Hankel-based integration and 4× faster than FFT-based synthesis. (b) Cross-correlation coefficients between our approximation and the FFT-based simulation. Our simulated PSF achieves high accuracy when the spherical aberration coefficient $C_s$ is less than 2.}
    \label{fig:time_n_accuracy}
\end{figure}

\paragraph*{Time and accuracy.} Fig.~\ref{fig:time_n_accuracy} compares the runtime and accuracy among Hankel, FFT, and the proposed approach (ours). We exclude Geometric, as it is not a wave-based PSF simulator. The proposed method achieves a clear reduction in computational cost relative to the baselines while accurately reproducing defocus and spherical aberration effects. It should be noted that the baseline FFT utilizes highly optimized Scipy functions, whereas our method currently uses a non-optimized Python script. The speed advantage of our approach would therefore become even greater with additional optimization.

\begin{figure}[t!]
    \centering
    \includegraphics[width=\linewidth]{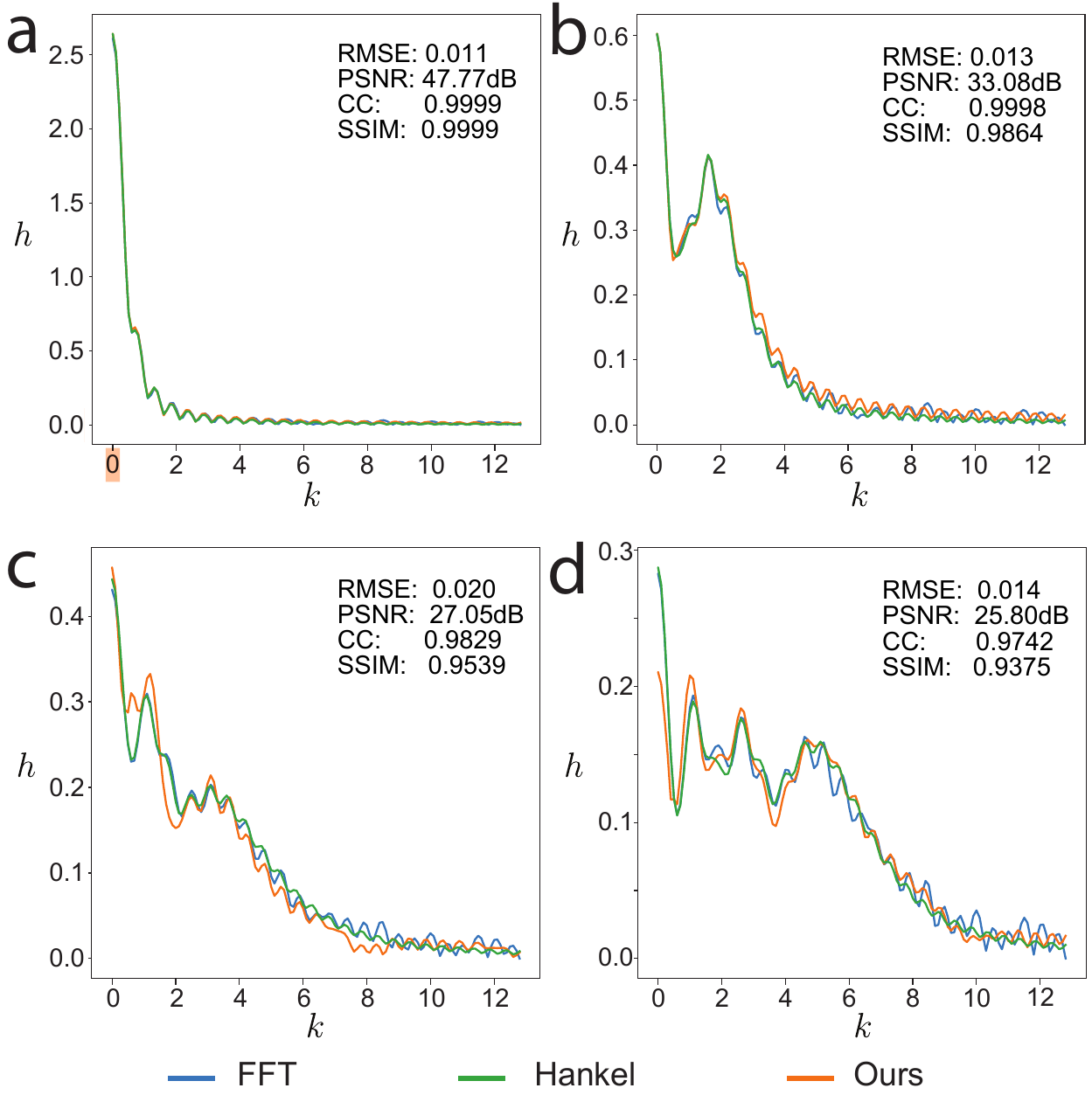}
    \caption{Comparison of radial PSF cross-sections using different simulators with sample defocus and spherical coefficients, $C_d$ and $C_s$. The quantitative metrics RMSE, PSNR (based on local maximum value), and correlation coefficient comparing our method to the baseline FFT approach are reported for the 1D amplitude profiles. SSIM is measured for the associated 2D PSF. Total energy deviation from the baseline is (a) 1.0\%, (b) 6.1\%, (c) 0.9\%, (d) 2.3\%. Parameter settings $[C_d, C_s]$: (a) $[1,0]$, (b) $[5,0]$, (c) $[5,5]$, (d) $[10,2]$. }
    \label{fig:psfs}
\end{figure}

Fig.~\ref{fig:psfs} further presents representative radial cross-sections of PSFs synthesized by different methods, demonstrating close qualitative agreement between our results and those of prior wave-based simulators when $C_s$ is small. The experiment is conducted on a machine with an Intel Core i7-11370H 3.30GHz, Quad-Core Processor and a 32GB RAM. 

\begin{figure*}
    \centering
    \includegraphics[width=\linewidth]{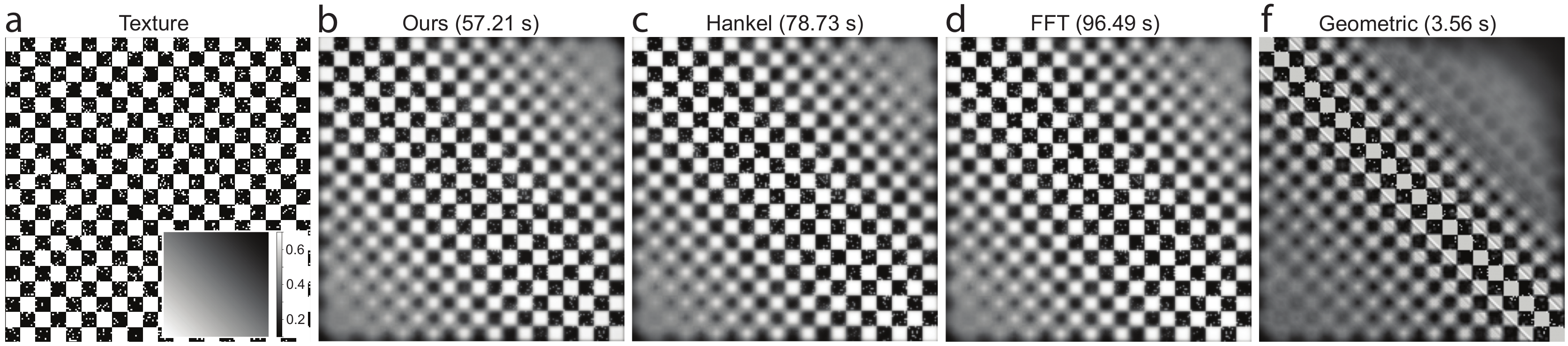}
    \caption{Sample depth-of-field images rendered using different simulators. We calculate the PSF of each pixel according to the provided depth map (inset, unit: m) and sum the PSF weighted by the brightness of the pixel. The focal distance is $z_f = 0.4$ m, the sensor distance is $s = 12.37$~mm, the pupil radius is $R = 1$~mm, and the wavelength is $\lambda = 500$ nm.  Ours achieves a clear speed advantage compared to Hankel and FFT. Although Geometric achieves the fastest rendering speed, it creates clear artifacts when the image is close to focus due to the discretization error of the scaling model.}
    \label{fig:dof}
\end{figure*}

\paragraph*{Depth-of-field (DoF) rendering.} Fig.~\ref{fig:dof} demonstrates a potential use case of the proposed PSF simulator. To accurately render the DoF effect in an image, we need to densely synthesize the PSFs of the scene. Our method's speed advantage leads to faster rendering than Hankel or FFT (Fig.~\ref{fig:dof}b-e). Although Geometric can lead to even faster rendering, the rendered image suffers from discretization artifacts when the target texture is close to focus, as shown in Fig.~\ref{fig:dof}f.

\section{Discussion}
\label{sec:discussion}

The proposed approximate closed-form PSF evaluation is complementary to the classic Hankel or FFT-based PSF evaluation. Ours clearly shows an advantage in computational time, but it can only handle defocus and spherical aberrations at the current stage. The latter are more suitable for rendering PSFs for custom pupil functions; in constrast, ours is better suited for synthesizing large-scale, realistic DoF images for depth from defocus, image deblurring, etc., as well as potentially serving as a computationally efficient method of obtaining synthetic defocused PSF training data for deep learning. On the other hand, the classic Hankel or FFT-based PSF simulators are superior for end-to-end design of computational imaging systems.

\section{Appendix}
\label{sec:appendix}
\paragraph*{Full derivation of the closed-form solution.} Eq.~\ref{eq:full_integrand} can be decomposed into the weighted summation of the following six forms of integrals:
\begin{equation}
\begin{aligned}
h_1(k) &\triangleq \int_{0}^{a_0(k)} r\,\Phi(r)\,dr, \\[4pt]
h_2(k) &\triangleq \int_{0}^{a_0(k)} r^{3}\,\Phi(r)\,dr, \\[4pt]
h_3(k) &\triangleq \int_{0}^{a_0(k)} r^{5}\,\Phi(r)\,dr, \\[6pt]
h_4(k) &\triangleq \int_{a_0(k)}^{R} C(r,k)\,\Phi(r;k)\,dr, \\[4pt]
h_5(k) &\triangleq \int_{a_0(k)}^{R} r\,C(r,k)\,\Phi(r;k)\,dr, \\[4pt]
h_6(k) &\triangleq \int_{a_0(k)}^{R} r^{2}\,C(r,k)\,\Phi(r;k)\,dr.
\end{aligned}
\end{equation}
where 
\begin{equation}
    \begin{aligned}
a_0(k) &\triangleq 1/(2\pi k), \\[4pt]
\Phi(r) &\triangleq \exp\!\left(j\,2C_d r^2/R^2\right), \\[4pt]
C(r,k) &\triangleq \cos\!\left(2\pi k r - \frac{\pi}{4}\right).
\end{aligned}
\end{equation}
The weighted summation is:
\begin{align}
    h(r) \approx 4\pi^2 \left|\sum_{i=1}^6 c_i(k) h_i(k)\right|^2,
\end{align}
where the coefficients are:
\begin{equation*}
    \begin{aligned}
c_1(k) &= 1, \quad
c_2(k) = -\pi^2 k^2, \quad 
c_3(k) = \frac{\pi^4 k^4}{4}, \\[6pt]
c_4(k) &= \sqrt{\frac{2}{\pi}}\;\frac{1}{2\pi k}, \quad 
c_5(k) = \sqrt{\frac{2}{\pi}}\;\frac{3}{2}\alpha^{-1/2}, \\
c_6(k) &= -\sqrt{\frac{2}{\pi}}\;\pi k\,\alpha^{-3/2}.
\end{aligned}
\end{equation*}
The integrals $h_{1-6}$ admit closed-form analytical solutions.
For $h_{1-3}$, by applying the change of variables $u=r^2$,
each integral reduces to the form $\int u^n e^{j(2C_d)u}\,du$,
which can be evaluated in closed form using elementary functions.
For $h_{4-6}$, the cosine modulation can be expressed as the sum of two complex exponentials:
$C(r)=\frac{1}{2}\{e^{-j\pi/4}e^{j2\pi k r} + e^{j\pi/4}e^{-j2\pi k r\}}$.
Each integral can then be written as the real part of an oscillatory integral
of the form $\int r^m e^{j(\nu r^2+\gamma r)}\,dr$ with
$\nu=2C_d$ and $|\gamma|=2\pi k$.
Completing the square in the phase converts the $m=0$ case into a complex
Gaussian integral, yielding a closed-form expression in terms of the complex
error function.
The cases $m=1$ and $m=2$ follow from algebraic recurrence identities that
express higher-order moments in terms of the base integral and boundary
exponential terms. The closed-form solutions are as follows. We omit the variable $k$ for $h_{1-6}$ and $a_0$ for simplicity.
\begin{equation}
\small
\begin{aligned}
h_1 &= \frac{e^{j2C_d a_0^2}-1}{4jC_d}, \\[6pt]
h_2 &= e^{j2C_d a_0^2}\!\left(\frac{a_0^2}{4jC_d}+\frac{1}{8C_d^2}\right)-\frac{1}{8C_d^2}, \\[6pt]
h_3 &= e^{j2C_d a_0^2}\!\left(\frac{a_0^4}{4jC_d}+\frac{a_0^2}{4C_d^2}-\frac{1}{8jC_d^3}\right)+\frac{1}{8jC_d^3}, \\
h_4 &= \frac{e^{-j\pi/4}}{2}\Big(F_0(R,k)-F_0(a_0,k)\Big) \\
&\qquad + \frac{e^{+j\pi/4}}{2}\Big(F_0(R,-k)-F_0(a_0,-k)\Big)   
, \\[6pt]
h_5 &= \frac{e^{-j\pi/4}}{2}\Big(F_1(R,k)-F_1(a_0,k)\Big)\\
&\qquad+ \frac{e^{+j\pi/4}}{2}\Big(F_1(R,-k)-F_1(a_0,-k)\Big), \\[6pt]
h_6 &= \frac{e^{-j\pi/4}}{2}\Big(F_2(R,k)-F_2(a_0,k)\Big)\\
&\qquad+ \frac{e^{+j\pi/4}}{2}\Big(F_2(R,-k)-F_2(a_0,-k)\Big).
\end{aligned}
\end{equation}
The intermediate elements $F_{0-2}$ has the forms:
\begin{equation}
\small
    \begin{aligned}
F_0(r,k) &= \mathcal{K}(r,k), \\[6pt]
F_1(r,k) &= \frac{e^{j(2C_d r^2+2\pi k r)}}{4jC_d}
        - \frac{2\pi k}{4C_d}\,\mathcal{K}(r,k), \\[8pt]
F_2(r,k) &= \frac{(4C_d r-2\pi k)e^{j(2C_d r^2+2\pi k r)}}{16jC_d^2} \\
        &\quad - \left(\frac{1}{4jC_d}-\frac{(2\pi k)^2}{16C_d^2}\right)\mathcal{K}(r,k),
\end{aligned}
\end{equation}
where
\begin{equation*}
\small
    \begin{aligned}
\mathcal{K}&(r,k)
\;\triangleq\;
\exp\!\left(-j\frac{(2\pi k)^2}{8C_d}\right)
\frac{\sqrt{\pi}}{2\sqrt{j\,2C_d}}\,
\mathrm{erf}\!\left(
j\,\sqrt{j\,2C_d}\left(r+\frac{2\pi k}{4C_d}\right)
\right). 
\end{aligned}
\end{equation*}

\newpage 
\small
\bibliographystyle{IEEEbib}
\bibliography{strings,refs}

\end{document}